\documentclass[superscriptaddress,groupedaddress,nofootnoteinbib,12pt]{article}  

\usepackage{amsmath}
\usepackage{amsfonts}
\usepackage{authblk}
\usepackage{bbm}
\usepackage{color}
\usepackage[dvipsnames]{xcolor}
\usepackage{graphicx,graphics}
\usepackage{epstopdf}
\usepackage{caption}
\usepackage{subcaption}
\usepackage{float} 
\usepackage{pifont}
\usepackage{titlesec}
\usepackage{etoolbox}
\usepackage{indentfirst}
\usepackage{jcappub}
\usepackage{braket}
\usepackage{ulem}
\usepackage{dsfont}
\usepackage{slashed}
\usepackage{tikz}
\usepackage{cancel}

\usepackage{mciteplus}

\usepackage{upgreek}

\usepackage{bookmark,hyperref}
\hypersetup{
	colorlinks,
	citecolor=blue,
	filecolor=blue,
	linkcolor=blue,
	urlcolor=blue
}

\def\bb{\begin{equation}}
\def\ee{\end{equation}}
\def\ba{\begin{array}}
\def\ea{\end{array}}
\def\babc{\begin{subequations}}
\def\eabc{\end{subequations}}

\def\5{\hspace*{5mm}}
\def\2{{\scriptstyle\frac12}}

\def\M{\bar M}
\def\s{\sigma}
\def\g{\bar g}
\def\e{\epsilon}

\newcommand{\beq}{\begin{equation}}
\newcommand{\eeq}{\end{equation}}

\newcommand\redsout{\bgroup\markoverwith{\textcolor{red}{\rule[0.5ex]{2pt}{0.4pt}}}\ULon}

\tikzset{every node/.style={align=center}}

\mciteSetSublistMode{s}
\mciteSetSublistLabelBeginEnd{\item[]}{}{}

\begin{document}


\vspace{0.5in}

\begin{center}

\Large{\bf Weakly-Coupled Trace Anomaly Action for Gravity}

\vspace{0.5in}

\large{Gregory Gabadadze\,$^{a}$ and Giorgi Tukhashvili$^b$}

\vspace{0.2in}

\large{\it $^a$Center for Cosmology and Particle Physics, Department of Physics, New York University, 726 Broadway, New York, NY, 10003, USA \\
\it $^b$Department of Physics, Princeton University, Princeton, NJ, 08544, USA }

\vspace{0.3in}

\end{center}

We discuss a local, diff-invariant quantum effective action for gravity 
that captures the trace anomaly  via a counter-term. We discuss why this counter-term is most significant among infinitely many possible ones, and show how the counter-term leads to a scattering amplitude that is strongly coupled at arbitrarily low energies. We show how the introduction of a new sector with  spontaneously broken scale invariance removes the strong coupling problem, and discuss some physical consequences due to the new sector. Three Appendices summarize quantum effective actions -- highlighting connections between their local, and seemingly non-local formulations -- for the scale anomaly in 4D QED, for the axial anomaly in 2D QED, and for the scale anomaly in a 2D sigma model.

\vspace{2in}
\begin{center}

    {\it To appear in the Valery Rubakov Memorial Volume}
\end{center}

\newpage


\section{Introduction}\label{into_sum}

The goal of this paper is to study an effective action 
for a gravitational field coupled to a quantum field theory 
(QFT) of spin-0,-1/2,-1 particles (e.g., the Standard Model of particle physics). The emphasis is going to be on 
the trace anomaly (a.k.a. scale anomaly) in the stress-tensor 
of a QFT in the gravitational field 
\cite {Capper:1975ig,Duff:1977ay}, in particular, on the question 
of how the effective  action captures this anomaly \cite{Riegert:1984kt,Fradkin:1983tg,Komargodski:2011vj,Gabadadze:2020tvt}. 

This question is of a certain importance: it is more convenient  
-- especially in gravitation and cosmology -- to capture the 
effects of the anomaly via the variation of the quantum effective action, rather than to start with solutions of the classical action and then calculate quantum anomaly  modifications of these solutions.

Recent investigations of this question lead to an unexpected 
conclusion that either the trace anomaly should be canceled in the 
full theory,  or else there has to exist a new sector 
responsible for spontaneous breaking of 
the scale symmetry with an associated spin-0 
Nambu-Goldstone boson \cite{Gabadadze:2023quw,Gabadadze:2023tgi,Gabadadze:2024wgj}. 

The requirement of the anomaly cancellation is well known  
for the anomalous symmetries that are gauged, but is not
usually imposed on a global symmetry, such as the scale invariance 
discussed in this work. This cancellation would
require spin-3/2 states, or more exotic tensor 
states \cite {Fradkin:1983tg}. 

It is equally unusual for 
an anomaly of a global symmetry to necessitate 
the  introduction of a new 
sector responsible for spontaneous breaking of the 
symmetry at hand. Indeed, global symmetries can in general be anomalous without being spontaneously broken. Nevertheless, without the introduction of the sector with spontaneous symmetry breaking, the full theory is inconsistent in the present case.  
We will discuss the origin and consequences of this claim 
in the remainder of the paper.

Let us recall that the global 
scale transformations can be parametrized  by a coordinate 
independent quantity $\gamma$ as follows:
\beq
x^\mu \to e^{\gamma} x^\mu\,,~~\chi \to e^{-\gamma\cdot {\rm dim}_\chi} \chi\,,~~g_{\mu\nu} \to g_{\mu\nu}\,,
\label{stransform}
\eeq
where $x^\mu$ are space-time coordinates, $g_{\mu\nu}$ is a metric, and $\chi$ is a spin-0,-1/2,-1, 
field of canonical dimensionality ${\rm dim}_\chi$. The latter equals 
to 1 for both canonical scalars and vectors, and equals to 3/2 
for a spin-1/2 field in 4D. 

Furthermore, global diffeomorphism transformations read 
\beq
x^\mu \to e^{\beta} x^\mu\,,~~\chi \to e^{-\beta\cdot {\rm diff}_\chi} \chi\,,~~g_{\mu\nu} \to e^{-2\beta}\,g_{\mu\nu}\,,
\label{gdiffs}
\eeq
where ${\rm diff}_\chi$ is the diffeomorphism  
weight of a field $\chi$. It equals to 0 for a scalar, to 1 for a vector, to 2 for a tensor, etc. By performing the above two transformations,
(\ref {stransform}) and (\ref {gdiffs}) simultaneously, and choosing $\beta =-\gamma$, we get
\beq
x^\mu \to  x^\mu\,,~~\chi \to e^{-\gamma\cdot ( {\rm dim}_\chi - {\rm diff}_\chi)} \chi\,,~~g_{\mu\nu} \to e^{2\gamma}\,g_{\mu\nu}\,.
\label{scale}
\eeq
For simplicity, and following many other past works, 
we will refer (\ref {scale}) as the scale 
transformations, even though (\ref {scale}) is a result of the   
simultaneous scale and diff transformations. This is a justified 
convention since we will always preserve diffeomorphisms. 

A canonical scalar in 4D transforms under (\ref {scale}),  
but a canonical vector does not. The conventional 
scalar kinetic term is not invariant, 
while the Maxwell term is invariant under 
(\ref {scale}) in 4D.
In a scale invariant theory without a gravitational field,  
the corresponding conserved current is a scale (dilatation) current,
related to the respective 
stress-tensor as follows: $j^d_\mu =x^\nu T_{\mu\nu}$.
The conservation of the dilatation current implies that the trace of the stress-tensor is zero, $T^\mu_\mu=0$. 

In a generic QFT the trace of the stress-tensor is not zero. Furthermore, general relativity (GR) coupled to a QFT  violates  
the scale invariance  due to the Planck mass, and the trace has to satisfy the equation,  $T^\mu_\mu = - 2 M^2 R$,  where $M$ is proportional to the Planck mass, $M=M_{Pl}/\sqrt{2}$, and $R$ is the Ricci scalar. 

The essence of the trace anomaly is that  in order to preserve diff invariance in the full quantum theory, the trace of the stress-tensor has to acquire additional specific terms
\beq
T^\mu_\mu = T^\mu_\mu|_{classical} - a E + c W^2\,,
\label{trace}
\eeq
where $a$  and $c$ are coefficients and $E$, $W^2$ are 
certain combinations of quadratic curvature invariants, 
$Rim^2,Ric^2,R^2$ (see precise definitions in the next section).
\footnote{For simplicity, we left out the gauge field trace anomaly 
in (\ref {trace}), which will be included in the subsequent 
sections; we have also left out the $\square R$ term,
which has an arbitrary coefficient and is irrelevant for our 
discussions, see a footnote in Section \ref{sec_2}.}

The main question of interest in this paper is this: 
what is the quantum effective action that would  
give rise to (\ref {trace}) by using  the variational principle \cite {Riegert:1984kt,Fradkin:1983tg}?

Answering this question leads to the local and diff invariant 
Riegert action \cite {Riegert:1984kt,Fradkin:1983tg,Komargodski:2011vj,Gabadadze:2020tvt}, which 
requires further modification \cite {Gabadadze:2023quw}, reviewed  
in this paper.

In Section \ref{sec_2} we will show how the counter-term that is 
necessary to remove quadratic divergences, and to reproduce 
the finite trace anomaly equation (\ref {trace}),  
necessarily leads to problems of consistency of the 
quantum effective action at  a scale that is much smaller than 
the Planck scale. This problem does not manifest itself 
in conventional applications of the gravitational theory with quantized fields, nevertheless, it is a consistency problem that should not be present in any effective field theory (EFT).  

This, and related problems are further 
discussed in Section \ref{sec_3}.  

Section \ref{sec_4} outlines how these problems 
are resolved by the introduction of a new sector responsible for spontaneous breaking of scale symmetry with an accompanying 
spin-0 particle \cite {Gabadadze:2023quw,Gabadadze:2023tgi},   
referenced as an anomalyon \cite{Gabadadze:2024wgj}.

In three Appendices we summarize quantum effective actions for other theories.
In particular, we discuss the  connection between the  local 
and seemingly non-local formulations of the effective actions 
for the scale anomaly in 4D QED (Appendix \ref{app_A}), for the axial anomaly in 2D QED (Appendix \ref{app_B}), and for the scale anomaly in a 2D Polyakov model (Appendix \ref{app_C}). While the contents of the last two Appendices are all well known and our presentation is of a review character, the discussion of the connection between the local and seemingly non-local formulations 
of the QED scale anomaly in gravitational field is 
novel, and it clarifies some confusions in the literature. In the paper we will be using the $(+,+,+)$ conventions from \cite{Misner:1973prb}.

Before we turn to the next section we'd like to make comments on the literature. A number of works adopt  what's called  
the Riegert non-local action, given in eq. (24) of the same paper \cite {Riegert:1984kt} where Riegert's  local action was derived. 
The Riegert's non-local action - eq. (24) of \cite {Riegert:1984kt} -- 
physically differs from his local action given in his eq. (8). 
The latter was also derived at the same time by Efim Fradkin and Tseyltin \cite{Fradkin:1983tg}. The non-local action, that is (24) of \cite {Riegert:1984kt},
could be made local by ``integrating in" a scalar degree of freedom, but  
at the expense of a four-derivative kinetic term for the scalar, as given 
in eq. (25) of \cite {Riegert:1984kt}. This four-derivative kinetic term 
gives rise to a ghost. The ghost is problematic, it leads to uncontrollable instabilities, both classical and quantum (see 
e.g., \cite {Cline:2003gs}, and many works  
before and after).  Yet, this action,  eq. (25) of 
\cite {Riegert:1984kt},  is often discussed in the literature. 
In our view, this ghostly action is not physical  in any 
Lorentz-invariant causal approach, and should be abandoned. 
In this paper we will be discussing only 
the Riegert's local cation, that is eq. (8) of \cite {Riegert:1984kt},
which is the same as the Fradkin-Tseytlin action \cite {Fradkin:1983tg}, and is
the action re-derived in \cite {Komargodski:2011vj},  and later in \cite{Gabadadze:2020tvt}, as an EFT without a four-derivative 
scalar kinetic term. 




\section{Trace anomaly from a counter-term }\label{sec_2}

For simplicity, let us begin with  an action of GR coupled to  
the Maxwell, Dirac, and Klein-Gordon 
massless  fields, written in conventional notations:  
\beq
S = \int d^4 x \sqrt{g} \left ( M^2 R   -{1 \over 4 e^2} 
F_{\mu \nu} F^{\mu \nu} + {\bar \psi} i \hat D \psi -  {1\over 2} 
(\partial \varphi)^2 \right )\,.
\label{RenGRQED}
\eeq
We'd like to quantize the Maxwell, Dirac, and Klein-Gordon fields and obtain 
an effective action for the dynamical gravitational field. The relevant action is then the renormalised effective 
action:
\beq
S_{Ren} = S +S_{ct}\,,
\label{RenS}
\eeq
where $S_{ct}$ contains the counter-terms that remove the divergencies of 
the respective Feynman diagrams. The gauge field $A$ and scalar $\varphi$  may in general have  
classical backgrounds too, and if so
one would quantize fluctuations in a self-consistent manner \cite{Coleman:1973jx}. 
In what follows we will  focus on  
a simple case where the classical backgrounds of these fields 
are trivial, $A_{cl} = \varphi_{cl}=0$. 

It is convenient to use dimensional regularization, by continuing the theory to  
$D=4 -2\epsilon$ space-time dimensions.  The corresponding one-loop vacuum polarization 
Feynman diagrams -- with the Maxwell, Dirac, and Klein-Gordon fields running 
in the loops to which two external gravitational lines are attached --  
have long been calculated.  
These diagrams are divergent, and counter-terms have to be introduced to 
cancel the divergencies \cite{Capper:1975ig}. 
 The key counter-term reads as follows \cite{Duff:1977ay}:
\beq
S_{ct}= \mu^{D-4} \frac{ \Gamma(2-D/2) }{2} \int d^Dx \sqrt{g_D} \Big( a \, E_D - c \, \Omega_D 
+ \frac{ \beta_0^{QED} }{2} F_{\mu \nu} F^{\mu \nu}
+\Big) \cdots \,,
\label{SC}
\eeq
where $E_D= R_{\mu \nu \rho \sigma} R^{\mu \nu \rho \sigma}- 4 R_{\mu \nu} R^{\mu \nu} +R^2$ is the Euler density, $\Omega_D = R_{\mu \nu \rho \sigma} R^{\mu \nu \rho \sigma} -2 R_{\mu \nu} R^{\mu \nu} +R^2/3$, while 
the coefficients $a$ and $c$ depend on the field content, and 
are both nonzero in general. In particular, 
$a= (N_s +11N_f+ 62 N_g)/360 (4 \pi)^2$, and $c =(N_s + 6 N_f + 12 N_g)/ 120 (4\pi)^2$, 
with  the number of real scalars $N_s$, Dirac fermions $N_f$,
and gauge bosons $N_g$, all contributing with positive signs. $\beta_0^{QED} = N_f/ (12 \pi^2)$ is the first coefficient of the QED beta function.  For  brevity of presentation,  we have not included in (\ref {SC})  the counter-terms for the spin-1/2 and spin-0 wavefunction renormalization, 
but they will be included  below.\footnote{Calculations by Capper and Duff \cite{Capper:1975ig,Duff:1977ay} 
were done at one loop with two external gravitational lines, 
leading to a correct counter-term in the linearized theory. 
The counter-term above is a unique diff-invariant completion of the linear 
order counter-term, that also gives the correct trace anomaly equation \cite {Duff:1977ay}.  Note that $\Omega_D$ reduces to the Weyl tensor square only when  $D=4$.} 

Hence, the explicit form of the dimensionally-regularized and 
renormalized gravitational action is
\begin{align}\label{GRdimreg}
  \nonumber S_{Ren} = & \int d^D x \sqrt{g_D}\, \left ( M^{2-2\epsilon} R_D   -{1 \over 4 e^2} F_{\mu \nu} F^{\mu \nu} + {\bar \psi_0} i \hat D_0 \psi_0 -  {1\over 2} (\partial \varphi_0)^2 \right ) \\
  {} & + \mu^{-2\epsilon}\, \frac{\Gamma (\epsilon) }{2} \,\int d^Dx\,\sqrt {g_D}  \Big( a \, E_D(g) - c \, \Omega_D + \frac{ \beta_0^{QED} }{2} F_{\mu \nu} F^{\mu \nu}\Big)\cdots ~,
\end{align}
where we've written the fermion and scalar kinetic terms,  as well as  
the covariant derivative, in terms of the bare quantities, $\psi_0,\varphi_0, {\hat D}_0$. The above are the counter-terms that are required if only the spin-0,-1/2,-1, fields are quantized.\footnote{In addition, one could add the $\sqrt{g}R^2$ counter-term in the action 
with an arbitrary coefficient, which would then modify the anomaly equation
(\ref {trace}) by an additive  $\square R$ term with the respective arbitrary coefficient, but this term is arbitrary and of no interest here.}.

Quantization of gravity as an effective field theory would further 
require an infinite number of higher dimensional curvature invariants, 
e.g., $Rim^2 R, R^3, Ric^2 R^2$, $R\square R$, etc. These higher dimensional terms are suppressed by the respective higher powers of the 
Planck scale $M$, and as such, they cannot help the 
problems that emerge at very low energies due to the terms 
in (\ref{GRdimreg}). Hence, we will ignore these higher 
dimensional terms for now, although they will in general be 
included in an  EFT (see discussions in 
Section \ref{sec_4}). 

Let us look at the equations of motion that follow from (\ref {GRdimreg}). 
In particular, focus on the equation obtained by requiring the 
trace of the variation of the action to vanish 
\beq
\,g_D^{\alpha \beta}{2 \over \sqrt {-g_D}} \,{\delta  \over \delta g_D^{\alpha\beta} }
 S_{Ren}  =0\,.
\label{var0}
\eeq
This gives the relation:  
\beq
2 M^2 R = a E(g) - c \Omega (g) + \frac{ \beta_0^{QED} }{2} F_{\mu \nu} F^{\mu \nu} +{\cal O}( D-4)\, \cdots ~ ,
\label{var}
\eeq
where the terms which are not related to the scale anomaly have been omitted in (\ref {var}).

The first two terms on the right hand side of (\ref {var}) 
constitute the trace anomaly, which is finite, 
and comes entirely from the variation of the counter-term in 
(\ref {GRdimreg}). The thirds term is the gauge field part of the trace anomaly action \cite{Crewther:1972kn,Chanowitz:1972vd,Chanowitz:1972da,Adler:1976zt}; its derivation in dimensional regularization 
is summarized in Appendix \ref{app_A}. 

The action (\ref {GRdimreg}), however, has a problem. To exhibit it clearly, we consider a scattering amplitude  
of four specific D-dimensional graviton states described by (\ref {GRdimreg}). 
For our purposes, it is sufficient to calculate parts of this amplitude 
that depend on $a$,  and are independent of $c$. This is so because 
$a$ and $c$ have different scaling with $N_s$, $N_f$, and $N_g$, and  cancellations 
between the terms containing only $a$, and those containing $c$, 
is not possible for generic values of $N_s,N_f$, and $N_g$.  Therefore, in 
what follows we will focus on the amplitude determined by the $a$-term only.

The calculation of such an amplitude has been done  in \cite{Bonifacio:2020vbk} in 
a different context, but the results of \cite{Bonifacio:2020vbk} are directly applicable 
to the dimensionally regularized action (\ref{GRdimreg}).  
Following \cite{Bonifacio:2020vbk}, we choose the polarization tensor of a D-dimensional 
graviton state to be 
\beq
\varepsilon^{AB} = {1\over 2} \left ( {1\over 2 } +{1\over D-4} \right )^{-1/2} 
\left ( \varepsilon^{\mu\nu}\delta_\mu^A \delta_\nu^B - {2\over (D-4)} 
\delta^{mn} \delta_m^A \delta_n^B \right )\,,
\label{Dpolarisation}
\eeq
where $\mu,\nu = 0,1,2,3,$ are four dimensional indices, and $m,n = 4,5,6,...D-1$, 
while the $4\times 4$ polarization matrix is given by
\beq
\varepsilon^{\mu\nu} =
\begin{pmatrix}
0 & 0 & 0 &  0\\
 0 & 1 & 0 &  0\\ 
 0 & 0 & 1 &  0\\ 
 0 & 0 & 0 &  0\\
    \end{pmatrix}.
    \eeq
The polarization tensor $\varepsilon^{AB}$ is traceless, 
is properly normalized, and is transverse to the momentum of a particle moving in the third direction of the conventional three-dimensional space, $p^A = (E,0,0,E, 0,...,0)$. 
The four-particle scattering amplitude, each with the above polarization, were shown 
to contain the following terms \cite{Bonifacio:2020vbk}:
\bb\label{ampl}
{\cal A} = \frac{s^2 +t^2 +u^2}{2 (D-4) M^4} \,r 
- \frac{3 s t u}{2 (D-4) M^6} \,r^2 + {\cal O} 
\left ( (D-4)^0\right )\,,
\ee
where $s,t,u,$ are the Mandelstam variables, $r \equiv a \,\Gamma (2-D/2) \, (\mu / M)^{D-4} $, 
and we omitted terms that are finite when $D\to 4$. 

The first two terms on the right hand side of (\ref {ampl}) 
diverge in the $D\to 4$ limit,  even if $r$ were to be finite. Of course, 
$r$ itself is a divergent coefficient of the counter-term in (\ref {GRdimreg}),
but the $1/(D-4)$ terms shown in (\ref {ampl})
constitute additional divergencies on top of the one contained 
in $r$.

To understand the origin of these additional divergencies we recall 
that the counter-term was introduced to cancel the divergencies 
in one-loop QFT diagrams with two external gravitational lines \cite{Capper:1975ig,Duff:1977ay,Christensen:1978gi}. 
Strictly speaking,  such a counter-term is quadratic in 
an external gravitation field. 
One needs to complete  this quadratic counter-term to a full fledged 
nonlinear term. In general this completion is not unique. 

Nevertheless,  there 
are two requirements for this completion that remove the arbitrariness \cite{Capper:1975ig,Duff:1977ay,Christensen:1978gi}: 
(1) the counter-term has to be fully diffeomorphism invariant, and (2) the variation of this
counter-term should give the correct trace anomaly equation in the $D\to 4$ limit. 
As soon as these two conditions are applied, the counter-term written in (\ref {GRdimreg}) is 
uniquely specified \cite {Duff:1977ay}. 

It is then the  nonlinear interactions introduced by this 
counter-term that unavoidably give rise to the additional divergencies in the 
four particle  scattering amplitude (\ref {ampl}). To reiterate,  
we've  started with one-loop two-external-line diagrams, but it is the diff-invariant nonlinear completion of the counter-term for these diagrams that gives rise to strongly coupled behavior. In terms of the Feynman diagrams, the
particular amplitude considered above corresponds to the ones with four external lines.

Note that these divergent terms can't be 
canceled by the corrections to the same 
four-particle  amplitude coming from the 
one-loop diagrams with the Maxwell, Dirac, and Klein-Gordon fields 
running in the loops, and with the four external lines attached to them. 
This is so because the latter are diverging as 
$1/\epsilon$, while the  amplitude (\ref {ampl}) contains 
stronger divergencies proportional to $r/(D-4)$ and $r^2/(D-4)$.

Moreover, higher loop diagrams with four external gravitational 
lines can have stronger than $1/\e$ divergencies, but they come 
with higher powers of the respective coupling constants, and therefore 
can't cancel the divergencies in (\ref {ampl}).

The root cause of this additional divergence is the fact that 
the renormalised D-dimensional action (\ref {GRdimreg}) contains additional degrees 
of freedom. One of these degrees of freedom is a 4D scalar appearing as a component of the 
D-dimensional graviton described by the polarization tensor (\ref {Dpolarisation}). 
This scalar becomes infinitely strongly coupled in the $D\to 4$ limit,
leading to  the additional divergencies seen in (\ref {ampl}). 
In the next section we will show explicitly  how  this scalar interacts in 4D, and why it is infinitely strongly coupled. 


\section{The local Riegert action and its shortcomings}\label{sec_3}

The Lagrangian in the action functional (\ref {GRdimreg}) 
is local when it's  written in terms of the $D$-dimensional 
metric tensor. However, this does not guarantee that (\ref {GRdimreg}) has a 
local 4D counterpart that can solely be written in terms of the 4D metric tensor. 
If the 4D action is 
required to be written only in terms of the 4D metric, then it will be nonlocal. 
Instead, a {\it local} 4D action can  be written in terms of the 4D metric 
{\it and} a scalar field \cite{Riegert:1984kt,Fradkin:1983tg}.

Let us see how this comes  about.   It is subtle to 
take the $\epsilon \to 0$ limit 
in (\ref {GRdimreg}), since the coefficient 
$\Gamma(2-D/2)$ is proportional to  
$1/\epsilon $, while the integrand
contains terms that scale linearly with $\epsilon$, and the ones 
that are independent of $\epsilon$. As a result, $S_{ct}$ contains a term 
which is finite in the $\epsilon \to 0$ limit, as well as the one that diverges 
as $1/\epsilon$. 
The former should give a local diff-invariant 4D action that reproduces the trace anomaly equation in $D=4$ (\ref {var}), and the latter will remove the one-loop divergencies of the bare action.  

One way to obtain the correct 4D action dictated by the symmetries and anomalies  
is to  assume the validity of a continuation  to small  
negative values of $\epsilon$.  Indeed,  by representing the  interval of  the 
$4+n$ dimensional theory as, 
\beq
d_{4+n}s = e^{2\sigma} (\g_{\mu\nu}(x) dx^\mu dx^\nu + d^2_n z)\,,
\label{dimreginterval}
\eeq
where $n=-2\epsilon^{-}$, and  then taking the limit $n=-2 \epsilon^{-} \to 0$,  the divergent coefficient proportional to $1/\epsilon^{-}$ coming from the $\Gamma$ function will be multiplied by a finite term,  and a term proportional to $\epsilon ^-$, both coming from the counter-term. The former will cancel with the one loop
$1/\epsilon^-$ divergencies coming from the bare action, while the latter will be 
proportional to $\epsilon^- \times (1/\epsilon^-)=1$, and independent 
on the sign of $\epsilon$. 

The  resulting finite term was worked out in a different context in \cite{Lu:2020iav} (see, also  \cite{Bonifacio:2020vbk} and references therein), and reads
\beq
a\int d^4x \sqrt {{\g} } \Big( \sigma { \bar E}  - 4 {\bar G}^{\mu\nu} {{\bar \nabla}}_\mu  \sigma {{\bar \nabla}}_\nu  \sigma  - 
4 ({{\bar \nabla}}^2  \sigma ) ({{\bar \nabla}}  \sigma)^2 - 
2 ({ {\bar \nabla}}  \sigma )^4 \Big) \,.
\label{Riegert_gsigma}
\eeq
This is what's called  the local Riegert action, initially 
derived in a different approach in 
\cite {Riegert:1984kt,Fradkin:1983tg}, an re-derived for other   
purposes in \cite {Komargodski:2011vj},  and in \cite {Gabadadze:2020tvt} (see below). It correctly  captures the 
a-term of the trace anomaly in the equations of motion. 
 
 As to the term with $\Omega_D$, it can be rewritten as follows
\beq
\mu^{D-4} \Gamma(2-D/2) \,c \,\int d^Dx \sqrt{g_D} \Omega_D = \mu^{D-4} \Gamma(2-D/2) \,c \,\int d^Dx \sqrt{\g_D}e^{(D-4)\s} {\bar \Omega}+ \cdots \,, 
\label{Omega}
\eeq
and contains a divergent part proportional to $1/\epsilon^-$, 
but also a finite part 
emanating from 
$\Gamma(2-D/2) \,c\, (D-4)\s {\bar \Omega}$ which in the $D=4$ 
limit gives a finite term proportional to $c\s W_{\mu \nu \rho \sigma} W^{\mu \nu \rho \sigma}$.

Hence the finite part of the total effective gravitational action is 
\begin{align}\label{RenGRbar}
  \nonumber S_{Ren} = &  \int d^4 x \sqrt{\g} \Big( M^2e^{-2\s} ({\bar R} +6 (\partial\s)^2) \Big) + c \int d^4x \sqrt {{ \g}} \sigma { \bar W}^{\mu \nu \alpha \beta} { \bar W}_{\mu \nu \alpha \beta} \\
  {} & - a \int d^4x \sqrt {{\g} } \Big( \sigma { \bar E}  - 4 {\bar G}^{\mu\nu} {{\bar \nabla}}_\mu  \sigma {{\bar \nabla}}_\nu  \sigma  - 
4 ({{\bar \nabla}}^2  \sigma ) ({{\bar \nabla}}  \sigma)^2 - 
2 ({ {\bar \nabla}}  \sigma )^4 \Big) ~ \cdots ~,
\end{align}
where dots stand for the parts of the counter-term with divergent 
coefficients proportional to $1/\epsilon^-$, as well as for the other 
fields present in (\ref {GRdimreg}). None of the omitted terms affect the 
trace anomaly. 

The first term on the r.h.s. is the EH action written in the Jordan frame. The second line is the Riegert action. Variation of the above action w.r.t. $\sigma$, with a subsequent substitution of ${\bar g}_{\mu\nu} = e^{-2\s} g_{\mu\nu}$, gives the anomaly equation (\ref {var}), as was first shown in \cite{Riegert:1984kt}. 

The above action  can be rewritten in terms of the metric $g$, bringing 
it to the Einstein frame:
\begin{align}\label{RenGR}
  \nonumber S_{Ren} = & \int d^4 x \sqrt{g}  M^2 {R} +c \int d^4x \sqrt {{ g}} \sigma { W}^{\mu \nu \alpha \beta} W_{\mu \nu \alpha \beta} \\
  {} & - a \int d^4x \sqrt {{g} } \Big( \sigma { E}  + 4 {G}^{\mu\nu} {{\nabla}}_\mu  \sigma {{\nabla}}_\nu  \sigma  - 
4 ({{\nabla}}^2  \sigma ) ({{\nabla}}  \sigma)^2 + 
2 ({ {\nabla}}  \sigma )^4 \Big) \, \cdots ~.
\end{align}
This very action has independently been reconstructed
by Komargodski and Schwimmer \cite{Komargodski:2011vj} (see, also an earlier work, \cite {Schwimmer:2010za})
as a unique  local 4D action reproducing the correct trace anomaly 
equation.  Furthermore, the above action can be obtained as the Wess-Zumino (WZ) term of the $SO(1,4)/ISO(1,3)$ coset \cite{Gabadadze:2020tvt}. 
As a WZ term,  this action can be represented 
as a boundary term  of a 5D action \cite{Gabadadze:2020tvt}. 
Interestingly, the analogous construction in two dimensions with the 
$SO(1,2)/ISO(1,1)$ coset, gives exactly the Polyakov action  \cite{Gabadadze:2020tvt}. 

It is clear that (\ref {RenGR}) is a potentially problematic 
effective field theory because the scalar $\sigma$ has nonlinear 
interaction terms, but has no quadratic terms. This fact has a multiple  
negative manifestations, three of which we will discuss in detail below. 

i) Let us start with the four-particle amplitude discussed in the 
previous section (\ref {ampl}). In the 4D language the amplitude 
corresponds to that of $2\to 2$  scattering of $\sigma$ particles 
in the flat 4D Minkowski background. In order to be able
to describe $\sigma$ states we regularize naively the theory by 
introducing a vanishing  kinetic terms  for $\sigma$ 
\beq
- u^2 M^2 \sqrt {g} (\partial\sigma)^2, 
\label{sd}
\eeq
where a dimensionless parameter $u$ is taken to zero 
at the end of the calculations.  It is instructive to consider this amplitude in the limit, $M\to \infty$, $u \to 0$, while  $ u \cdot M \to 0$.
Then, the  leading term in the forward $2\to 2$ 
scattering amplitude for $\sigma$  particles comes from the $(\nabla \sigma)^4$ 
term in (\ref {RenGR})  and reads
\beq
{\cal A}_{forward} \propto \, {a\,s^2 \over (u\cdot M)^4}\,, 
\label{ampf}
\eeq
which  shows that the effective strong scale for the interactions 
of the $\sigma$ particles is 
\beq
{u\, M \over a^{1/4} }\to 0\,.
\label{strongs}
\eeq
This is a 4D manifestation of the 
problem of the infinite strong coupling discussed in 
the previous section.

ii) In the absence of a quadratic term for $\sigma$  in (\ref {RenGR}),
the $\sigma$ kinetic term would be entirely determined by the classical solutions for the metric and $\sigma$, both induced by an external classical source $T_{\mu\nu}$ coupled to $g$. For simple backgrounds with weak gravitational  
field and weak background, $\sigma_b<<1$, the kinetic term  
for the fluctuations $\delta \s = \s-\s_b$ will be determined by 
the respective curvatures of the gravitational background, and by derivatives of the $\sigma$ background, as it is clear from (\ref {RenGR}).  
Such background-dependent kinetic terms would in 
general lead to instabilities and/or super-luminal propagation. 
For instance, for a weak stress-tensor coupled to the metric,   
$G_{\mu\nu} \simeq 8\pi G_NT_{\mu\nu}$, and  the 
kinetic term  for a perturbation $\delta \sigma$ will receive  a contribution proportional to $T^{\mu\nu} {\partial}_\mu \delta \sigma { \partial}_\nu \delta \sigma$. For any source which has positive pressure (e.g., a planet) the spatial part of such a kinetic term  has unhealthy sign, and would 
lead to gradient instabilities of $\delta \sigma$. These instabilities
could be avoided if the kinetic terms emerging from the 
other three terms in (\ref {RenGR}) could flip the signs of the 
gradient terms, but this could happen only for particular sources,  
and not in general. 

iii) The equation of motion for $\sigma$ derived from (\ref {RenGR}) reads:
\begin{align}\label{SEQ}
  \nonumber {} & a \Big( { E} - 8 {G}^{\mu\nu} {\nabla}_\mu { \nabla}_\nu \sigma  
+ 8 (({ \nabla}^2 \sigma)^2 - (\nabla_\mu\nabla_\nu \sigma)^2) - 
8 R^{\mu \nu} {\nabla}_\mu \sigma { \nabla}_\nu \sigma - 8
 \nabla^\mu ({\nabla}_\mu \sigma  ({\nabla}\sigma )^2 ) \Big) \\
  = & c ~ { W}^{\mu \nu \alpha \beta} W_{\mu \nu \alpha \beta}.
\end{align}
Consider the Minkowski space, with $\sigma =0$, 
which is a solution. 
In the quadratic order around the Minkowski background 
the above equation reads: 
\beq
a \Big( {{{ E}^{(2)} }} - 8 {G^{(1)}}_{\mu\nu} {\partial }^\mu { \partial}^\nu \delta\sigma  - 8 (({ \partial}^2 \delta \sigma)^2 - (\partial _\mu\partial_\nu \delta \sigma)^2) \Big) = c  { W^{(1)} }_{\mu \nu \alpha \beta} { W^{(1)} }^{\mu \nu \alpha \beta} \,. 
\label{SEQ2}
\eeq
Here ${ E}^{(2)} \sim (\partial \partial h^{(1)})^2$ is the Euler density 
calculated to a quadratic order for the first order metric perturbations $h^{(1)}$, 
and $G^{(1)}$ and $W^{(1)}$ denote, respectively, the Einstein and Weyl tensors in the linearized approximation for the first order metric perturbations.  Eq. (\ref {SEQ2}) imposes a  constraint on the first order perturbations, instead of determining the second order perturbations, as it would ordinarily be 
the case in a systematic nonlinear tree-level perturbation theory in GR.  Indeed, all first order solutions have to satisfy the  second order equation (\ref {SEQ2}). For instance, a linearized Schwarzschild solution 
that has a nonzero Kretschmann scalar, and therefore nonzero $E^{(2)}$, would 
excite  a nonzero $\delta \s$ that would have to satisfy (\ref {SEQ2}).  
While this is not necessarily a  fatal problem by itself, 
it is a serious complication which is absent in GR.


\section{The anomalyon kinetic term}\label{sec_4}

The discussions in the previous two sections show that the trace anomaly 
makes it problematic for GR to couple to quantized spin-0,-1/2,-1 fields 
(this applies to GR coupled to the Standard Model of particle physics). 

On the other hand, the problem  manifests itself neither in classical GR calculations nor in calculations of quantized perturbations in the early universe. These calculations are not sensitive to the nonlinear terms in the trace anomaly expression\footnote{The exception could be 
the Starobinsky inflationary model \cite{Starobinsky:1980te}, or its generalizations, where the  issue with the $\s$ kinetic term 
would show up  if both the tensor and scalar perturbations were considered.
That said, on cosmological backgrounds $\s$ would acquire a background depended kinetic term,  with accompanying problems, as discussed in the previous section.}. 
Nonetheless, the theory (\ref {RenGR}) is not a meaningful EFT,  
and the problem needs to be solved. 

One way to address the problem is to consider the theories in which the 
trace anomaly could cancel to all orders. As we discussed, 
spin-0, spin-1/2 and spin-1
fields  all contribute with positive sign to the values of both 
$a$ and $c$ and can't lead to the trace anomaly cancellation. However, a spin-3/2 field is known to contribute with a negative sign,  and one can identify specific  
super-multiplets in certain supergravities where such cancellations  can  
take place, at least in the one-loop approximation \cite{Fradkin:1983tg,Birrell:1982ix}. 
This cancellation would  be a strong argument in favor of the existence of spin-3/2 
gravitini, and perhaps also some other higher rank exotic tensor 
states  \cite{Fradkin:1983tg,Birrell:1982ix}.  Such an approach would be somewhat 
similar to the one Nature chose in the Standard Model of particle physics 
where the axial anomaly is canceled in each generation of fermions, although there is a significant difference too: the trace  anomaly is an anomaly in a global symmetry, while the axial anomaly that is canceled in each Standard Model generation is for the local (gauged) symmetry.  

Short of the anomaly cancellation for global scale symmetry, 
a way to deal with the above problem 
is to postulate the existence of a quadratic kinetic term for $\sigma$. 
This however is not as trivial as it sounds, one can't just add a finite  
kinetic term for $\s$ to (\ref {RenGRQED}) because 
it would modify the anomaly equation (\ref {var}) by an additional term. 
This would give  a ``wrong" trace anomaly equation, 
not supported by GR coupled 
to the spin-0,-1/2,-1 fields. 

There is however a more subtle way to introduce the $\s$ kinetic term 
without introducing extra terms in the trace anomaly equation \cite{Gabadadze:2023quw}. 
It is easier to understand this mechanism in the Jordan frame,  in terms of the metric  
$\bar g$ and $\sigma$. If we add to (\ref {RenGRQED}) a new term:
\beq
{\bar S}(\g) = - \M^2 \int d^4x \sqrt{\g} R(\g) = - 
\M^2 \int d^4x  \sqrt{g} \Big(e^{-2\s} ({R} +6(\partial\s)^2) \Big) ,
\label{GRbar0}
\eeq
we would get the sigma kinetic term with the right sign, and at the same 
time the variation of (\ref {GRbar0}) would not modify the 
trace-anomaly equation \cite{Gabadadze:2023quw} (we emphasize that the sign of the Einstein-Hilbert (EH) term in (\ref {GRbar0}) has to be opposite to the sign  of the conventional EH term  in order for the conformal scalar hidden in (\ref {GRbar0})  
to have the right-sign kinetic  term). The dimensionful parameter $\bar M$
is supposed to be much smaller than the Planck scale, $M$. This smallness guarantees that the wrong-sign tensor kinetic term that emanates from (\ref {GRbar0}) is smaller than the right-sign kinetic term coming from the standard Einstien-Hilbert action.  

The existence of the new term, (\ref {GRbar0}), would signify the existence of a new sector in the gravitational theory, 
with $\sigma$  being a Nambu-Goldstone boson of spontaneously 
broken scale invariance at the scale $\bar M$. This breaking is in addition to the explicit breakings introduced by the Planck mass,  and by all the other dimensionful parameters of the Standard Model.

To summarize, the  local weakly-coupled 
diffeomorphism invariant effective action 
for the trace anomaly  has the 
following form \cite{Gabadadze:2023quw,Fernandes:2021dsb}
\beq
S_{tot} = S(g)+ {\bar S}(\g)+ {\bar S}_{A}(\g,\sigma)\,,
\label{actiontotal}
\eeq
with the two metrics related as,  $\g_{\mu\nu} = e^{-2\s} g_{\mu\nu}$, 
the actions $S(g)$  and  $\bar S({\bar g})$  defined respectively 
in (\ref {RenGRQED}), and  (\ref {GRbar0}), and  the anomaly  
action ${\bar S}_{A}(\g,\sigma)$ of \cite{Riegert:1984kt,Fradkin:1983tg} given by:
\beq
- a \int d^4 x\sqrt{\g} \left (  \sigma {\bar E} - 4 {\bar G}^{\mu \nu} {\bar \nabla}_\mu \sigma {\bar \nabla}_\nu \sigma 
	- 4 \left( {\bar \nabla} \sigma \right)^2 {\bar \nabla}^2 \sigma - 2 \left( {\bar \nabla} \sigma \right)^4   - \frac{c}{a} \sigma {\bar W}_{\mu \nu \alpha \beta} {\bar W}^{\mu \nu \alpha \beta} \right).
\label{Riegertbar}
\eeq
Here ${\bar E}$ denotes  the Gauss-Bonnet invariant, ${\bar R}_{\mu \nu \rho \sigma} {\bar R}^{\mu \nu \rho \sigma} - 
4 {\bar R}_{\mu \nu} {\bar R}^{\mu \nu} + {\bar R}^2$, while 
${\bar G}_{\mu \nu}$ and ${\bar W}_{\mu \nu \alpha \beta}$  are the Einstein and Weyl tensors, respectively.

The effective action of those massless  fields which define the values of $a$ and $c$ should in general be included in 
(\ref {actiontotal}). These fields might have their own classical backgrounds, and/or could have significant perturbations in a background of other fields, depending on a particular cosmological scenario. They are omitted here just for simplicity of presentation. 

Furthermore, the EH action $S(g)$ should be supplemented 
by higher dimensional curvature invariants of $g$, suppressed by $M$,
while the action $\bar S(\g)$ should be supplemented by higher
dimensional curvature invariants of $\g$, suppressed by $\M$ \cite {Gabadadze:2024wgj}.

In the limit when gravity decouples, 
$M\to \infty$, with $\M$ being fixed, the action (\ref {actiontotal})  
reduces to the action of the $\s$ field with a conformal kinetic term,  and 
non-linear Galileon interactions that are suppressed by the scale $\M$ \cite{Gabadadze:2023quw}:
\beq
{\cal L}_{Dec} = - 6 e^{-2\s/{\M}}(\partial \s)^2 
+a \left (  
{4 {{\square}}  \sigma  ({{\partial}}  \sigma)^2\over \M^3} - 
{2 ({ {\partial}}  \sigma )^4 \over \M^4 }\right )\,,
\label{dec}
\eeq
where we've rescaled the $\s$ field as $\s \to \s/\M$.  Using this approximation  one can easily  calculate the scaling of the forward scattering amplitude of the $2\to 2$ sigma particles :
\beq
{\cal A}_{Forward} \sim a {s^2 \over \M^4}\,.
\label{Afors}
\eeq
This amplitude entirely comes from the quartic Galileon term in (\ref {dec}) (possible contributions from the cubic Galineon vertices vanish on-shell). The higher dimensional curvature invariants 
of $g$ vanish in the decoupling limit and do not modify 
the amplitude (\ref {Afors}) at all, and the higher dimensional 
curvature invariants of $\g$ would modify  (\ref {Afors}) 
only in higher orders of the derivative expansion, i.e., by 
terms of ${\cal O}\left (  s^3/\M^6\right )$, which are subleading 
at low energies, $\sqrt {s}<< \M$.

If $\M$ is set to zero, as it is the case in the Riegert theory, 
the amplitude (\ref {Afors}) is divergent. This is consistent with the 
divergence  discussed in Section 3. For any nonzero $\M$, the 
above amplitude is weak for lower energies, and only 
becomes strong  at the momentum scale $\sim \M/a^{1/4}$.
In a realistic scenario of particle physics, we do not expect $a$  
to be too large, hence we will not differentiate the above 
scale from $\M$ (except when we consider the large $N$ limit, 
later in this section). 

Can the theory (\ref {actiontotal})  be extended above the strong scale $\M$? Asked differently, is there a weakly-coupled higher energy theory that would naturally give rise to  (\ref {actiontotal}) in the low energy approximation?  The affirmative action to this question was given 
in \cite{Gabadadze:2023tgi}, in the large $N$ approximation, where it was shown that 
the theory naturally emerges in a holographic Randall-Sundrum model 
\cite{Randall:1999ee}, with additional boundary terms providing sub-leading 
large $N$ corrections. This framework explains  
the existence of the sector with spontaneously broken scale symmetry at $\M$, and provides a geometric interpretation of the hierarchy between the scales of $M$ and $\M$, with $M>> \M$ \cite{Gabadadze:2023tgi}.  
Indeed, in this approach 
the Planck scale $M_{Pl}\sim M\sim N/L$, while $\M\sim N/z_c$,
where $L$ is the radius curvature of the 5D AdS spacetime, and $z_c$ is the distance between two branes in the 5th dimension, with $z_c>> L$. 

Given these relationships, it is instructive to work out  the scaling of the coefficients of various nonlinear terms in (\ref {actiontotal}) in the large $N$ limit. We list them below one-by-one for each nonlinear term as they appear in the action:
\beq
a \s E \rightarrow \Lambda^{-3}_{\s E}\sim {a\over \M M^2}\sim {z_c L^2\over N}\,, \nonumber \\
\eeq
\beq
a G \nabla\s\nabla\s \rightarrow \Lambda^{-3}_{G\s\s}\sim 
{a\over  M \M^2}\sim {L z_c^2\over N}\,, \nonumber \\
\eeq
\beq
a (\nabla^2 \s) (\nabla \s)^2 \rightarrow \Lambda^{-3}_{3}\sim {a\over \M^3 }\sim {z^3_c\over N}\,, \nonumber \\
\eeq
\beq
a (\nabla\s)^4  \rightarrow \Lambda^{-3}_{4}\sim {a\over \M^4 }\sim {z^4_c\over N^2}\,, \nonumber \\
\eeq
\beq
c \s W^2 \rightarrow \Lambda^{-3}_{\s W^2 }\sim {c\over \M M^2}\sim {z_c L^2\over N}\,. \nonumber \\
\label{Nscaling}
\eeq
The lowest scale on the above list is the one by which the cubic 
Galileon is suppressed, $\Lambda_3 \sim N^{1/3}/ z_c$. Fortunately, this scale is higher than the scale $1/z_c$,  which is the mass scale of the Kaluza-Klein excitations that are responsible for the softening of the amplitude already in the 4D regime; this soft behavior continues all the way up to the  
scale $1/L>> 1/z_c$. Above the scale of $1/L$ the theory is 
five-dimensional.

The sigma field is massless  during inflation \cite{Gabadadze:2024wgj}. 
This has interesting cosmological and astrophysical consequences \cite{Gabadadze:2024wgj}. If $\s$ couples to the matter fields via 
the gauge field trace anomaly, as it is the case in the present framework, then it will  acquires a thermal temperature-dependent mass 
after the reheating. Furthermore, it will acquire a 
non-perturbatively generated mass due to interactions 
with non-Abelian gauge fields (with QCD or 
some other higher energy confining theory) \cite{sigmamass}. 
Hence, the $\s$ field is massive in the present day universe, 
with the mass scale below $\M$ \cite{sigmamass}.


\section*{Acknowledgements}

We'd like to thank Kurt Hinterbichler, Massimo Porrati, David Spergel, and Arkady Tseytlin for useful communications during various stages of this work. GG is supported in part by the NSF grant PHY-2210349.


\appendix

\vskip 2cm

\section{Trace anomaly in QED}\label{app_A}

Consider a renormalized action for QED  with one  massless fermion in Minkowski space, written in terms of the 
bare coupling constant, $e_0$,  bare gauge field strength $F_0$, and bare fermion field 
$\psi_0$:
\beq
{\cal L}^{Ren} = -{1 \over 4 e_0^2} {F_0}_{\mu \nu} F_0^{\mu \nu} + {\bar \psi}_0 i \hat D_0 \psi_0\,.
\label{RenQED}
\eeq
We use dimensional regularization in $D=4 -2\epsilon$  
to express bare quantities in terms of the physical ones; for instance, in the one-loop approximation the bare coupling constant $e_0$ is related to the physical one $e$ as  
follows, $e_0^{-2} = e^{-2} - \beta_0^{QED} \e^{-1}$, where $\beta_0^{QED} = (12 \pi^2)^{-1}$ is the first coefficient of the QED beta function.

A convenient way to derive the trace anomaly is to introduce an  auxiliary  metric field 
in QED and take variation of the action with respect to that field. In the present case this is not only convenient but also physically relevant since in the full 
theory the metric is dynamical.  For now, and for clarity,  we regard the 
metric as lacking its own dynamics and introduce $g_{\mu \nu}=e^{2\sigma} {\g_{\mu \nu}}$ as a spurious feild in the QED action:
\beq
S^{Ren}_g = \int d^Dx \sqrt{\g} \left [ e^{ -2 \e \s} \left ( -{1 \over 4 e_0^2} {F_0}_{\mu \nu} F_0^{\mu \nu} \right )  + 
e^{ (3-2\e) \s}  {\bar \psi}_0 i \hat D_0 \psi_0   \right ]\,,
\label{QEDs1}
\eeq
where all the indices are contracted via $\g$. Furthermore, expressing the bare parameters in terms of the physical ones we get
\beq
S^{Ren}_g  = \int d^Dx \sqrt{\g} \left [ e^{ - 2\e \s} \left ( -{1 \over 4 e^2} F_{\mu \nu} F^{\mu \nu} + K_{e}  F_{\mu \nu} F^{\mu \nu} \right )  + 
e^{ (3- 2\e) \s}  \left ( {\bar \psi} i \hat D \psi + K_{\psi}  {\bar \psi} i \hat D \psi \right ) \right ]\,,
\label{QEDs2}
\eeq
where $K_e$ and $K_\psi$ are the divergent coefficients of the counter-terms containing both the charge and fields' renormalisations. Their finite parts  
are renormalization scheme dependent, but the divergent parts are not,  in particular $K_e =\beta_0^{QED} / (4 \epsilon) +finite~terms$.
Expanding the exponents and keeping  only those $\sigma$ dependent terms which are proportional to $\e K_e\sim O(1)$ and $\e K_{\psi}\sim O(1)$,  we get
\beq
S^{Ren}_g  = \int d^Dx \sqrt{\g} \left [ \left ( -{1 \over 4 e_0^2} {F_0}_{\mu \nu} F_0^{\mu \nu}  - 2\e K_{e} \s F_{\mu \nu} F^{\mu \nu} \right )  +  
 \left ( {\bar \psi}_0 i \hat D_0 \psi_0  - 2\e K_{\psi} \s {\bar \psi} i \hat D \psi \right ) +O(\e) \right ]\,,
\label{QEDs3}
\eeq
where we used both the bare and physical fields in the 
same expression, keeping in mind the relationships between them. 
Taking the variation of the latter action with respect to $\s$, 
and using the equations of motion, we get for the trace of the stress tensor
\beq
T^\mu_\mu = -2 \e K_e  F_{\mu \nu} F^{\mu \nu} = - \frac{\beta_0^{QED}}{2} F_{\mu \nu} F^{\mu \nu} \,.
\label{QEDanomaly}
\eeq
This is the well-know trace anomaly for a gauge field. 

\vskip 0.2cm

For a somewhat different  perspective on the trace anomaly, 
let us now derive the same in terms of the quantum effective action.
Since the path integral for fermions is Gaussian, the quantum 
effective action equals to
\beq
{S }_{eff}^{Ren} = \int d^D x  \left ( -{1 \over 4 e_0^2} {F_0}_{\mu \nu} F_0^{\mu \nu} - i Tr \, ln  (i \hat D_0 ) \right     )\,.
\label{RenQEDEff}
\eeq
The task is to calculate the gauge field dependence of the $iTrln (i \hat D_0 )$.  
In the one-loop 
approximation in dimensional regularization one gets 
\beq
i \, \int d^D x \, Tr\, ln  (i \hat D_0 ) \simeq \int d^D x  \, K_e {F_0}_{\mu \nu} \square^{-\e} F_0^{\mu \nu} \,+{\cdots}\,,
\label{Det}
\eeq
where the dots stand for the terms that are either independent of the gauge fields,
or are of higher order in them.
We now introduce spurious gravity. For this we use the following protocol: first we represent the
flat space operator  $\square^{-\e}$ as $(\square^2)^{-\e/2}$, and then introduce 
gravitational field in $\square^2$ as follows
\beq
\square^2 \to \Delta \equiv (g^{\mu \nu} \nabla_\mu \nabla_\nu)^2 + 2 R^{\mu \nu} \nabla_\mu \nabla_\mu - {2\over 3} R \nabla^2 + {1\over 3} (\nabla^\mu R)\nabla_\mu \,.
\label{Delta}
\eeq
This protocol is not unique, but the non-uniqueness is in 
terms that are higher order in the metric, and hence not important for us. The virtue of the Paneitz operator $\Delta$  is its homogeneous conformal transformation, $\Delta \to e^{-4\s}  {\bar \Delta}$,
under $g_{\mu \nu} \to e^{2\s} {\g}_{\mu \nu}$ \cite{Paneitz}.

Equipped with this trick we notice that the key term in (\ref {Det}) is conformally 
invariant in $D$-dimensions
\beq
\int d^Dx \sqrt{g} \  
{K_e} {F_0}_{\mu \nu} {\Delta}^{-\e/2} F_0^{\mu \nu} = \int d^Dx \sqrt{\g}  
{K_e} {F_0}_{\mu \nu} {\bar \Delta}^{-\e/2} F_0^{\mu \nu}  \,,
\label{QEDsEff}
\eeq
where on the l.h.s the indices are contracted by $g^{\mu \nu}$ and on the r.h.s. by $\g^{\mu \nu}$. Therefore, the quantum effective 
action (\ref {RenQEDEff}) can be rewritten as follows
\beq
{S }_{eff}^{Ren} = \int d^D x \sqrt{\g} \left [ e^{-2\e \s} \left (  -{1 \over 4 e_0^2} {F_0}_{\mu \nu} F_0^{\mu \nu} \right )
 -{K_e} {F_0}_{\mu \nu} {\bar \Delta}^{-\e/2} F_0^{\mu \nu}  \right  ]\,,
\label{RenQEDEffg}
\eeq
where all the indices are contracted via $\g$.
The last term on the r.h.s. in the above action does not  depend on $\s$, 
hence it does not contribute to the anomaly. 
The anomaly in the $\e \to 0 $ limit  
is given entirely by the divergent piece of 
$ e_0^{-2} {F_0}_{\mu \nu} F_0^{\mu \nu}$ multiplied by $-2\e \s$, and 
equals to (\ref {QEDanomaly}).

\vskip 0.2cm

There is yet another, seemingly different but factually equivalent 
perspective on the trace anomaly. To summarize it, we first expand (\ref {RenQEDEff})  and insert the expansion 
in the renormalized effective Lagrangian to get
\beq
{\cal L}^{Ren}_{eff} \simeq  -{1 \over 4 e^2} F_{\mu \nu} F^{\mu \nu} + {K_{e}  F_{\mu \nu} F^{\mu \nu}} - {K_e} F_{\mu \nu} F^{\mu \nu}   + \e K_e F_{\mu \nu} ln (\square)^{1/2} F^{\mu \nu} \,.
\label{log}
\eeq
In the above expression, the counter-term -- which is the second term on the r.h.s. -- cancels the divergent term coming from $Tr ln$, which is the third term on the r.h.s. 
After this cancellation, we introduce  gravity following the protocol specified above
by replacing $\square \to \Delta$,  
and  use the identity:  $ln \Delta^{1/2} = ln (e^{-2\s} {\bar \Delta}^{1/2}) = 
ln e^{-2\s} + ln {\bar \Delta}^{1/2}$, to get
\beq
S^{Ren}_{eff}  \simeq \int d^Dx \sqrt{\g} \left [  -{e^{-2\e \s} \over 4 e^2} F_{\mu \nu} F^{\mu \nu}  + 
\e {K_e} F_{\mu \nu} ln ( e^{-2\s}) F^{\mu \nu} +  \e {K_e} F_{\mu \nu} ln ({\bar \Delta})^{1/2}  F^{\mu \nu}   \right ]\,.
\label{QEDsEff2}
\eeq
The exponent $e^{-2\e\s}$ in front of the gauge kinetic term 
no longer matters since the divergent counter-term 
was canceled. Hence, the sigma dependence and the anomaly, 
arise from the non-local logarithmic term. However, in the  $\s$-$\g$ basis the sigma dependence of this nonlocal term 
becomes local, as clearly seen in the above expression. 
This is not a new anomaly, it is the same scale anomaly as 
before. 

Similar considerations can be applied to the gravitational 
trace anomaly. This clarifies the connection of the approach of  
\cite{Deser:1976yx} to the standard approach \cite{Capper:1975ig,Duff:1977ay,Christensen:1978gi}, and suggests  that 
the anomaly obtained in \cite{Deser:1976yx} is not a new anomaly, 
contrary to the claim of that paper, but is another  
way of obtaining the conventional  trace anomaly.


\section{The Schwinger model}\label{app_B}

We turn to the massless quantum electrodynamics in $(1+1)$ space-time dimensions, 
{\it a.k.a.} the massless Schwinger model. The action written in conventional notations is: 
\beq
{S}_{Sch} (A,\psi, \psi^+)=  \int d^2 x  
\left (  -{1 \over 4 e^2} F_{\mu \nu} F^{\mu \nu} + 
{\bar \psi} i \hat D \psi\,  \right     )\,.
\label{QED2}
\eeq
The above action is invariant under both the vector and axial global $U(1)$ transformations, 
resulting in the conservation of both the vector and axial currents. However, at the quantum level 
only one of these two conservation equations can be maintained. Requiring that the quantization 
respect the conservation of the current that couples the gauge field -- the vector current --  
one unavoidably gets an anomalous non-conservation of the axial current:
\beq
\partial^\mu ({\bar \psi} \gamma_\mu \gamma_5 \psi\ ) = {1\over 2\pi} \epsilon_{\mu\nu} F^{\mu\nu}\,.
\label{axial}
\eeq
Can we write an effective action  for the Schwinger model that would incorporate 
the quantum effects, and in particular give the anomaly  equation (\ref {axial})
as one of its equations of motion?

The answer to the above question is affirmative, as we summarize below. We will proceed by performing  
the Gaussian integral  over the fermion fields in a path integral with the action (\ref {QED2}) 
\beq
e^{i {S}_{eff} (A)}=\int \,d\psi^\dagger d\psi \,e^{i {S}_{Sch}(A,\psi, \psi^+)} \,,
\label{path2}
\eeq
where the effective action depends on the vector field only
\beq
{S }_ {eff}(A)= \int d^2 x  \left ( -{1 \over 4 e^2} F_{\mu \nu} F^{\mu \nu} - i \, Tr \, ln  (i\hat D) \right     )\,.
\label{QED2Det}
\eeq
The determinant of the Dirac operator can be calculated exactly in the massless Schwinger model \cite{Alvarez:1983th,Rothe:1985br}
\beq
- i \, ln\,det (i{\hat D}) = - i \, Tr \, ln (i \hat D) = - {1\over 2\pi}\,\,A_\mu \left ( \eta^{\mu\nu} - {\partial^\mu \partial^\nu \over \partial^2} \right )A_\nu\,.
\label{det}
\eeq
While the calculation of ref. \cite{Alvarez:1983th,Rothe:1985br} used the heat kernel approach, it is worth noting that 
the above exact answer  is given by a one-loop vacuum polarization diagram. Using  (\ref {det}) the total effective 
action can be written as  follows:
\beq
{S }_ {eff}(A)= \int d^2 x  \left ( -{1 \over 4 e^2} F_{\mu \nu} F^{\mu \nu}  + {1\over 4\pi}\, F_{\mu \nu} \,{1 \over \partial^2}\, F^{\mu \nu} \right     )\,.
\label{QED2NL}
\eeq
If one insists on keeping the effective action written in terms of the gauge field only, then 
the expression for the action is non-local. 

In reality, however, this apparent non-locality is
only  a consequence of the restriction to write the action solely in terms of the gauge field. 
One could easily ``integrate in" a scalar field \beq
  e^{i {S}_{eff} (A)} =    \int \,d\phi \,e^{i {S}_{eff} (A,\phi)}\,,
 \label{path22}
\eeq
to rewrite the action in a local form
\beq
{S }_ {eff}(A,\phi)= \int d^2 x  \left ( -{1 \over 4 e^2} F_{\mu \nu} F^{\mu \nu} - {1\over 2} (\partial \phi)^2 - {1\over \sqrt{4 \pi}}\,
\phi \epsilon_{\mu\nu}  F^{\mu\nu} \right     )\,.
\label{QED2Phi}
\eeq
The action (\ref {QED2Phi}) is the effective action containing all the quantum information. 
Let us see how this action accounts for the anomaly equation (\ref {axial}) via 
the equations of motion.  Variation of  (\ref {QED2Phi}) w.r.t. $\phi$ gives 
\beq
\square \phi = {1\over \sqrt{4 \pi}} \epsilon_{\mu\nu}  F^{\mu\nu} \,.
\label{anomaly}
\eeq
To show that this is indeed  equivalent to (\ref {axial}), we also look at the variation of  (\ref {QED2Phi}) 
w.r.t. the gauge field 
\beq
\partial^\mu F_{\mu\nu} = {e^2\over \sqrt{\pi}}\,\epsilon_{\nu\alpha} \partial^{\alpha} \phi. 
\label{Maxwell}
\eeq
From the latter it follows that the vector current is $J^V_\nu= \epsilon_{\nu\alpha} \partial^{\alpha} \phi/\sqrt{\pi}$; furthermore, 
in two dimensions the vector current and the axial current are related as $J^V_\nu = \epsilon_{\nu\beta}J^{A\beta}$,
hence the axial current is $J^A_\mu= \partial_{\mu} \phi/\sqrt{\pi}$. If so, then the left hand side of (\ref {anomaly})
equals to   $\sqrt{\pi}\,\partial^\mu J^A_\mu$, and  therefore eq. (\ref {anomaly}) is equivalent to eq. (\ref {axial}).  

The action (\ref {QED2Phi})  describes one massive scalar, which can be viewed as a Stueckelberg 
mode of a massive 2D vector field. This mode is a 2D 
analog of the massive anomalyon.


\section{The Polyakov action}\label{app_C}

Consider the following $(1+1)$ dimensional  
action 
\beq
S_2 (\gamma, \Phi) = \frac12 \int d^2 \zeta \,\sqrt {\gamma}\, \gamma^{\mu\nu} \partial_\mu \Phi^a \partial_\nu \Phi^b \eta_{ab}\,,
\label{Paction}
\eeq
where $\zeta_\mu\,$ denote two coordinates of a $(1+1)$  dimensional space-time 
endowed with the metric $\gamma_{\mu\nu},\,\mu,\nu=0,1$. There are 
$N$ scalars, $\Phi^a(\zeta)$, $a,b = 0,1,..., N-1$, propagating in this two dimensional spacetime, 
with the internal-space symmetry group of $ISO(1,N-1)$, and $\eta_{ab}$ is the $N$-dimensional 
Minkowski metric. The two dimensional Einstein-Hilbert term of the metric $\gamma$ is a surface term 
defining the Euler characteristic of the two-dimensional manifold and has been omitted in (\ref {Paction})  
for brevity, but its existence should be kept in mind. The very same action (\ref {Paction}) is well know 
to describe a bosonic string of unit tension propagating in a $N$-dimensional Minkowski spacetime  
endowed with the coordinates $\Phi^a$.

We'd like to look at an effective action $S_{2eff}$ defined by the following path integral
\beq
 e^{i {S}_{2eff}(\gamma)}=  \int [d\Phi] \, e^{i {S}_{2} (\gamma,\Phi)}\,.
\label{path3}
\eeq
Polyakov has reconstructed the  path integral (\ref {path3}) from the knowledge of the scale anomaly. 
In particular, the gauge  in which the metric  is conformally flat, $\gamma_{\mu\nu} = \rho (\zeta) \,\eta_{\mu\nu}$, leads 
to  the dependence of the classical action on the conformal factor $\rho$ to disappear. The Gaussian integral over $\Phi$ then leads 
to a determinant that's independent of $\rho$, but is divergent.  The only dependence on $\rho$ then arises via a 
counter term that has to be included to account for the correct conformal anomalous  equation.   
Hence, the resulting effective action -- that contains the counter-term --  
ought  to be \cite{Polyakov:1981rd}: 
\beq
S_{2eff} (\rho) = {N\over 48 \pi}\, \int d^2 \zeta \,\left ( \frac12 (\partial _\mu ln \rho )^2 +\mu^2  \rho \right )\,.
\label{Paction1}
\eeq
This can be rewritten in a gauge invariant but seemingly nonlocal form:
\beq
S_{2eff} (\gamma) = {N\over 48 \pi}\, \int d^2 \zeta \sqrt {\gamma} \,\left ( - \frac12 R(\gamma) {1\over \square} R(\gamma) +\mu^2 \right )\,.
\label{Paction2}
\eeq
Here $\mu^2$ is an induced 2-dimensional cosmological constant, and 
$\square\equiv (\gamma)^{-1/2}\,\partial_\mu \sqrt {\gamma} \gamma^{\mu\nu}\partial_\nu\,.$\,\footnote{Note that 
the further integration over quantum fluctuations of $\rho$ would give an additional factor of $-26$, making the 
overall coefficient in front of the anomaly term to be $N-26$, \cite{Polyakov:1981rd}.}
The apparent non-locality in (\ref {Paction2}) is however the consequence of the insistence on writing the 
action solely in terms of the metric $\gamma$. If this requirement is relaxed, the action can be written 
in a local form by integrating in a scalar field $\tau$:
\beq
S_{2eff}(\gamma,\,\tau) = {N\over 48 \pi}\, \int d^2 \zeta \sqrt {\gamma} \, \Big( -2 (\partial _\mu \tau  )^2 -2 \tau\, R(\gamma)  +\mu^2 \Big) \,.
\label{Paction3}
\eeq
The scalar $\tau$ has a healthy kinetic term. The above Polyakov action is a $2D$ analog of the  $4D$ weakly-coupled trace anomaly action in (\ref {actiontotal}).


\bibliographystyle{utphys_mciteplus}
\bibliography{refs}


\end{document}